\documentclass[aps,prx,twocolumn,showpacs,superscriptaddress,floatfix]{revtex4-1}

\usepackage{color}
\usepackage{amsmath,amssymb}
\usepackage{bm}
\usepackage{verbatim}
\usepackage{graphicx}
\usepackage{dcolumn}
\usepackage{amstext}
\usepackage{mathptmx}
\usepackage{verbatim}
\usepackage[colorlinks,citecolor=blue,bookmarks]{hyperref}
\usepackage{xcolor}
\usepackage{color,soul}
\def\veck{\mathbf{k}}

\def\dxz{c_{d_{xz}}}
\def\dyz{c_{d_{yz}}}
\def\pz{c_{p_{z}}}
\def\dXZ{d_{xz}}
\def\dYZ{d_{yz}}
\def\pZ{p_{z}}
\def\sx{\sigma_x}
\def\sy{\sigma_y}
\def\sz{\sigma_z}

\usepackage{subfigure}
\DeclareMathAlphabet{\mathpzc}{OT1}{pzc}{m}{it}

\begin{document}


\title{Band inversion and topology of the bulk electronic structure in FeSe$_{0.45}$Te$_{0.55}$}
%


\author{Himanshu Lohani}
\affiliation{Physics Department, Technion-Israel Institute of Technology, Haifa 32000, Israel.}
\author{Tamaghna Hazra}
\affiliation{Physics Department, Ohio State University, Columbus, OH 43210, USA}
\author{Amit Ribak}
\affiliation{Physics Department, Technion-Israel Institute of Technology, Haifa 32000, Israel.}
\author{Yuval Nitzav}
\affiliation{Physics Department, Technion-Israel Institute of Technology, Haifa 32000, Israel.}
\author{Huixia Fu}
\affiliation{ Department of Condensed Matter Physics, Weizmann Institute of Science, Rehovot 7610001, Israel }
\author{Binghai Yan}
\affiliation{ Department of Condensed Matter Physics, Weizmann Institute of Science, Rehovot 7610001, Israel }
\author{Mohit Randeria}
\affiliation{Physics Department, Ohio State University, Columbus, OH 43210, USA}
\author{Amit Kanigel}
\affiliation{Physics Department, Technion-Israel Institute of Technology, Haifa 32000, Israel.}



\date{\today}
\begin{abstract}
FeSe$_{0.45}$Te$_{0.55}$ (FeSeTe) has recently emerged as a promising candidate to host topological superconductivity,
with a Dirac surface state and signatures of Majorana bound states in vortex cores.
However, correlations strongly renormalize the bands compared to electronic structure calculations,
and there is no evidence for the expected bulk band inversion.
We present here a comprehensive angle resolved photoemission (ARPES) study of FeSeTe as
function of photon energies ranging from 15 - 100 eV. We find that although the top of bulk valence band
shows essentially no $k_z$ dispersion, its normalized intensity exhibits a periodic variation with $k_z$.
We show, using ARPES selection rules, that the intensity oscillation is a signature of band inversion 
indicating a change in the parity going from $\Gamma$ to Z. Thus we provide the first direct evidence for a
topologically non-trivial bulk band structure that supports protected surface states.
\end{abstract}


\maketitle
Iron-based superconductors (FeSCs) have been intensely investigated since their discovery in 2008~\cite{hsu} as strongly correlated
materials that harbor high temperature superconductivity. Recently, interest in this field has increased greatly due to new experiments 
that suggest that some of these systems may be topological superconductors~\cite{maj} that harbor Majorana bound states (MBS) in their vortex cores,
which could be potentially important for quantum information processing~\cite{Kit}.

Wang {\it et al.}~\cite{wang1} first suggested that FeSe$_{0.5}$Te$_{0.5}$ (FeSeTe) can host topologically protected Dirac surface states,
which were recently observed directly using angle resolved photoemission spectroscopy (ARPES)~\cite{ss}. Soon after, such states were found 
in other FeSCs~\cite{ss1} and in thin films~\cite{thin}. In addition, clear zero bias conductance peaks (ZBCP) were observed~\cite{stm,stm1} 
in the superconducting vortex cores in FeSeTe using scanning tunneling spectroscopy (STS), and identified as the MBS
expected in topological superconductors. In fact, the strong correlations in these materials, which leads to surprisingly large 
$\Delta/E_F$ ratios~\cite{kanigel1,kanigel}, helps in separating the ZBCP from (topologically) trivial vortex core bound states.

Despite these exciting developments, direct evidence for the topological nature of the {\it bulk} band structure -- 
responsible for the topologically protected surface states and MBS -- is lacking. 
Density functional theory (DFT) calculations~\cite{wang1} for FeSeTe find a $p_z$ band that is highly dispersive along 
$k_z$, which mixes with an appropriate linear combination of the $d_{xz,yz}$ bands. As a result, the orbital character and the parity 
of the band changes as one goes from $\Gamma$(0,0,0) to Z(0,0,$\pi$/c).
However, no such highly dispersive band is observed in the data, as we shall show below, and -- at first sight -- there seems to be no
evidence for the band inversion expected in a topologically nontrivial bulk band structure.

FeSeTe is known to be the most strongly correlated member of the FeSC family~\cite{tamai,kotl}, making it difficult to directly compare
ARPES measurements with DFT.
It offers an exciting opportunity to study the interplay between the topological nature of the band structure 
and the effect of the strong electronic correlations.  

In this letter, we present a systematic ARPES study of FeSeTe for a broad range of incident photon energies (15 to 100 eV) 
to investigate the $k_z$-dispersion of the bulk electronic structure. Using symmetry analysis and dipole selection rules, 
we present clear evidence for the change in the parity eigenvalue going from $\Gamma$ to Z, in spite of the absence of
any highly dispersive band. We also present a tight-binding model, with reasonable values of renormalization parameters relative to DFT 
and of spin-orbit coupling, which gives insight into ARPES observations. We thus provide compelling evidence for bulk ``band inversion'',
the hallmark of a topological band structure via the Fu-Kane invariant~\cite{fu},
which leads to a protected Dirac surface state in the energy gap near the $\Gamma$ point. 

\begin{figure}
\includegraphics[width=0.45\textwidth]{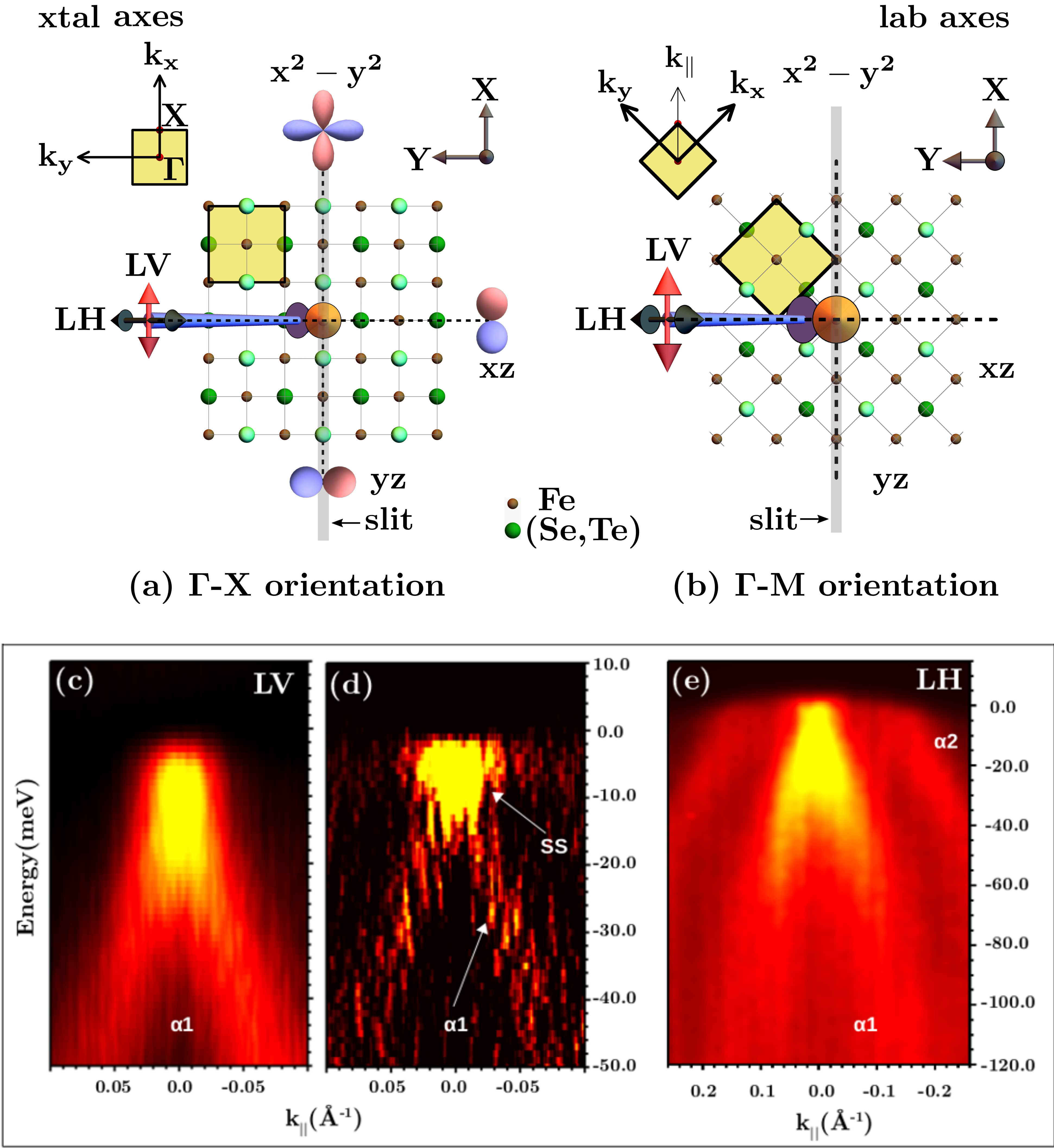}
\caption{\label{fig1} (a, b) Notation and conventions: s-type ARPES setup with 
the analyzer slit for emitted electrons in the XZ (vertical) plane, and light incident
in the YZ (horizontal) plane in either vertical (LV) or horizontal (LH) polarizations, as shown.
The sample is oriented with the analyzer slit along $\Gamma$-X (panel a) or $\Gamma$-M (panel b). 
The orbitals are  labelled with reference to the crystal coordinate axes $(x,y,z)$ while the laboratory
frame is denoted by $(X,Y,Z)$.
(c) ARPES image taken along $\Gamma$-M direction at T = 25 K using 22 eV LV polarized light.
 (d) Second derivative of the data shown in (c) to emphasize the Dirac surface state.
(e) ARPES image for the same sample measured using 22 eV LH polarized light.}
\end{figure}

We used high quality Fe$_{1.02}$Se$_{0.45}$Te$_{0.55}$ single crystals  for ARPES measurements. 
Fig.~\ref{fig1}(a,b) shows the geometry of our ARPES experiments.
We will focus on near-normal emission with $(k_x,k_y)$ near $(0,0)$, and light incident
in the YZ plane in either LV (linear vertical) or LH (linear horizontal) polarizations, as shown. This geometry will be crucial in the analysis of the
selection rules later in the paper. 
Our laboratory axes $(X,Y,Z)$ conform with the literature~\cite{ss,ss1},
however, we label orbitals with reference to the crystallographic axes $(x,y,z)$, irrespective 
of sample rotations, consistent with Refs.~\cite{chen,johnson,xu}.

We show ARPES data along the $\Gamma$-$ M $ direction using 22 eV LV photons in Fig.~\ref{fig1} (c), and its 
second derivative~\cite{cur} sharpened image in panel (d). This allows us to see in addition to a dispersive bulk band, which we label
as $\alpha_1$, an intense state at a binding energy (BE) of around 10  meV, that lies
between the top of the $\alpha_1$ band (BE $\simeq 18$ meV) and the chemical potential (BE $= 0$ meV).
This state is similar to the linearly dispersive Dirac surface state (SS),
recently been reported by  Zhang {\it et. al.}~\cite{ss}. 
In Fig.~\ref{fig1} (e) we show LH polarization data where in addition 
to the states seen in LV data of panel (c), we also see another dispersive $\alpha_2$ band.

The ARPES intensity allows a direct mapping of the 
electronic dispersion for momenta parallel to the sample surface. This is because only in-plane momentum is conserved
in the photoemission process. To map the dispersion along $k_z$ , one needs to scan as a function of the 
incident photon energy. We use the free-electron final-state approximation to find the correspondence between the photon energy and $k_z$; 
see Suppl. Info. Sec-I  for details.

\begin{figure}
\includegraphics[width=0.5\textwidth]{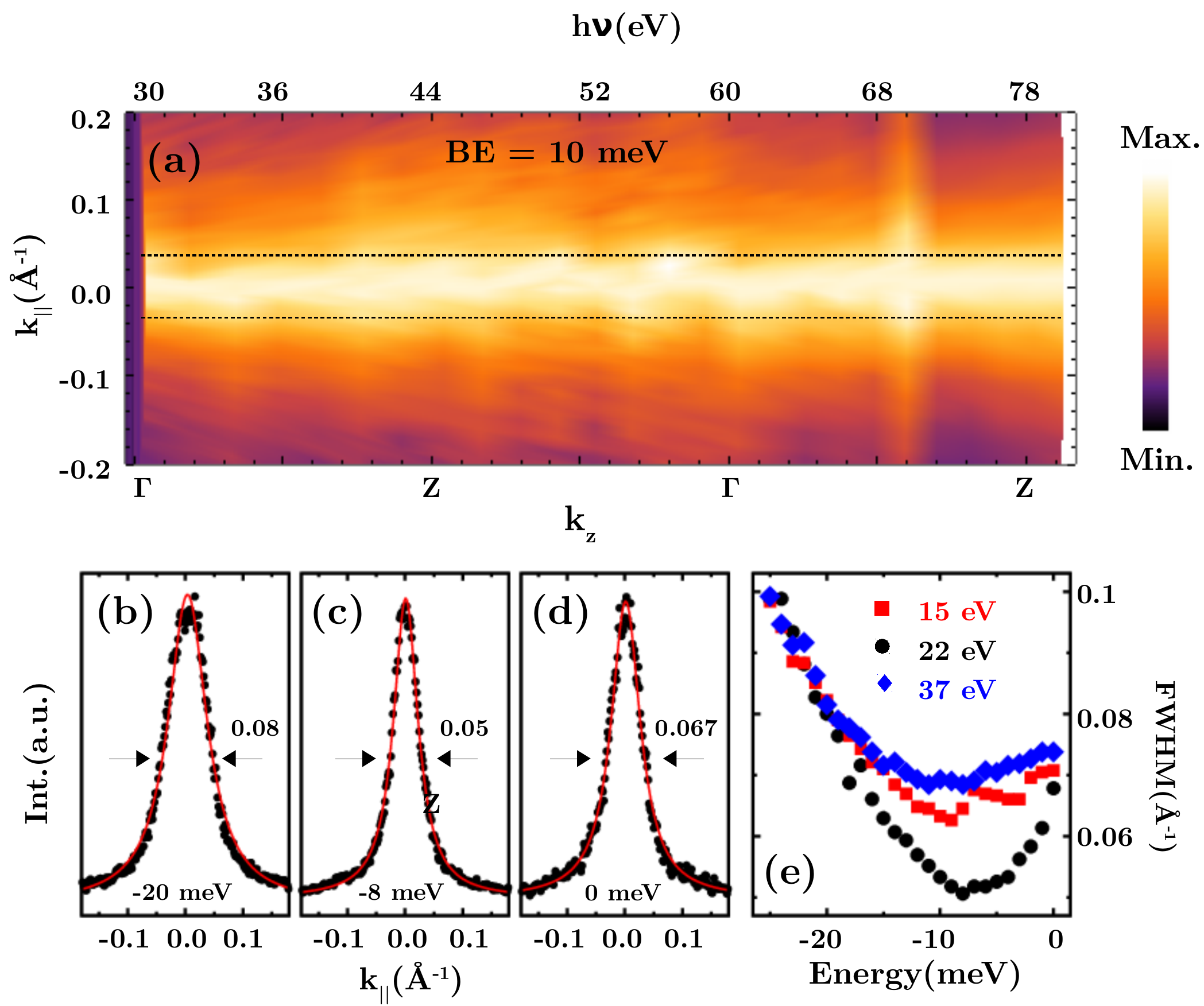}
\caption{\label{fig2} Surface state: (a) ARPES intensity map at binding energy (BE) of 10 meV, as a function of 
the in-plane momentum k$_\parallel$ along $\Gamma$-X (y-axis) and
photon energy, which probes different $k_z$ values (x-axis).
(b-d) MDCs at 20, 8 and 0  meV BE measured using 22 eV photons with LV polarization.The red lines are Lorentzian fits to the data.
(e) Full width at half maximum of Lorentzian fits as a function of BE, for three different photon energies. The BE of the minimum
in these curves gives an estimate of the location of the Dirac point.}
\end{figure}

Before turning to the bulk electronic structure ($\alpha_1$ and $\alpha_2$ bands), which is our main focus, we first look at
that intense state near 10  meV BE. In Fig.~\ref{fig2}(a), we show the
ARPES intensity map over an extensive photon energy range at a fixed BE $= 10$ meV,  where the x-axis represents $k_z$ (converted from photon energy)
and the y-axis represents k$_\parallel$ along $\Gamma$-X direction. 
We find intensity at this BE for all $k_z$ values, consistent with a surface state. 
At these photon energies the ${\bf k}$-resolution does
not allow us to extract the Dirac-like dispersion of the surface state, we can nevertheless estimate the
location of the Dirac point as follows. We make Lorentzian fits to the momentum distribution curves (MDCs) of the ARPES
intensity at a fixed BE as a function of $k_\parallel$ in Fig.~\ref{fig2} (b,c,d), and plot the full width at half maximum (FWHM) of the fits as
a function of BE in Fig.~\ref{fig2} (e). We estimate the Dirac point to be at 8  meV, the BE at which 
the MDCs exhibit a smallest FWHM. 
Crucially, the BE of the Dirac point is independent of the photon energy, as expected for a surface Dirac state.

We next turn to the band structure of the bulk $\alpha_1$ and $\alpha_2$ bands. We will discuss in detail below 
their orbital content, and the resulting constraints on ARPES selection rules. 
For now, suffice it to say that both are made up of Fe-derived $d_{xz},d_{yz}$ orbitals, 
and $\alpha_1$ also has an important $p_z$ admixture (See Appendix~\ref{sec:derivation}).

The in-plane dispersion of  $\alpha_2$ bands, shown in Fig.~\ref{fig1} (e), 
can be fit with a simple (hole-like) parabolic model to determine its top at $(k_x,k_y) = (0,0)$,
even when it lies above the chemical potential; see Appendix~\ref{sec:alpha2} for details.
The top of $\alpha_1$ band is obtained from the $(k_x,k_y) = (0,0)$ EDC peaks.
For $\alpha_1$ we use LV data (Fig.~\ref{fig1}(c)) and for $\alpha_2$ we use LH data (Fig.~\ref{fig1}(e))
and determine the tops of the bands as a function of $k_z$ by changing the incident photon energy.

\begin{figure}
\includegraphics[width=0.45\textwidth]{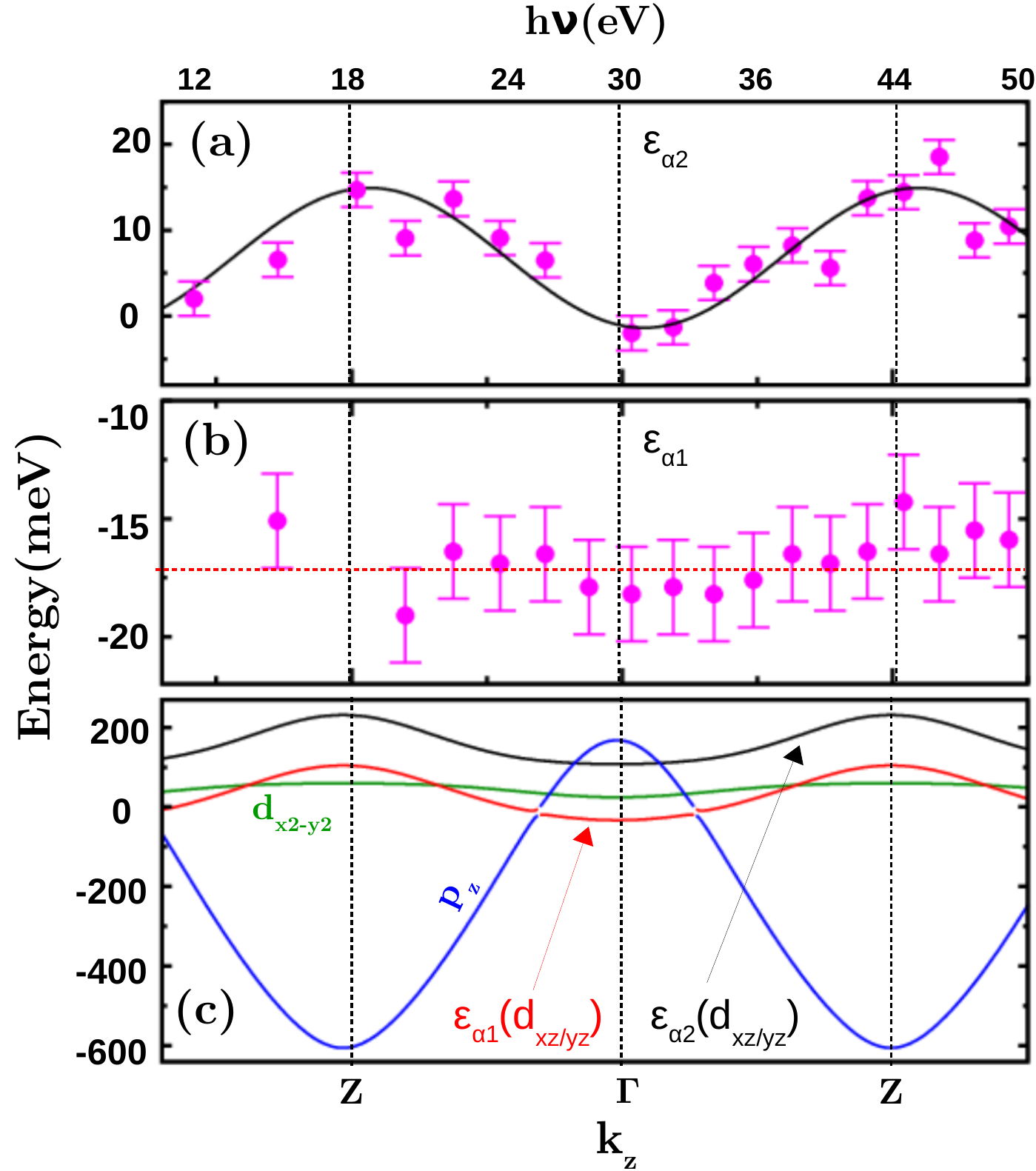}
\caption{\label{fig3} Bulk band structure: (a) $k_z$ dispersion of the $\alpha_2$ band 
extracted from LH polarization ARPES data at different photon energies. 
(b)  $k_z$ dispersion of the $\alpha_1$ band based on LV polarization ARPES data.
(c) DFT band structure along $k_z$ which is very different from the ARPES data; see text for details.}
\end{figure}

Our goal is to look for the band inversion predicted by DFT by mapping out the 
$k_z$-dispersion, going from $\Gamma$(0,0,0) to Z(0,0,$\pi$/c).
From Fig.~\ref{fig3} (a) we see that the top of the $\alpha_2$ band, $\epsilon_{\alpha_2}(0,0,k_z)$ 
shows a periodic variation with $k_z$ with a maximum at Z, 
a minimum at $\Gamma$, and a $k_z$-bandwidth of about 18  meV. 
In contrast, the corresponding result for $\epsilon_{\alpha_1}(0,0,k_z)$ in Fig.~\ref{fig3}(b) 
shows essentially no dispersion; see also Ref.~\cite{johnson}.

Let us compare these $k_z$-dispersions with the DFT results shown in Fig.~\ref{fig3}(c).
The observed $\alpha_2$ dispersion is at least crudely consistent with upper $d_{xz/yz}$ band in 
DFT, if one is willing to renormalize the bandwidth down by a factor of about 5 and make a shift in energy. 
However, the dispersion-less $\alpha_1$ band seems difficult to reconcile with 
100  meV wide $d_{xz/yz}$ band that crosses a 500  meV wide $p_z$ band in DFT.
The $d_{x^2-y^2}$ band is not seen in the experiments in the energy-momentum window that we focus on in this work.

To see how the ARPES data can be understood as a renormalized band structure 
with reasonable parameters, we turn to a tight binding model for the $k_z$ dispersion of FeSeTe.
This will also help us to see how selection rules can help address the question of the topological/trivial nature of the band structure.
We focus only on $(k_x,k_y) = (0,0)$ here, although one can use ${\bf k}\cdot{\bf p}$ perturbation theory
to look at $k_\parallel$ dispersion; see Appendix~\ref{sec:derivation}.

\begin{figure}
\includegraphics[width=7.5cm]{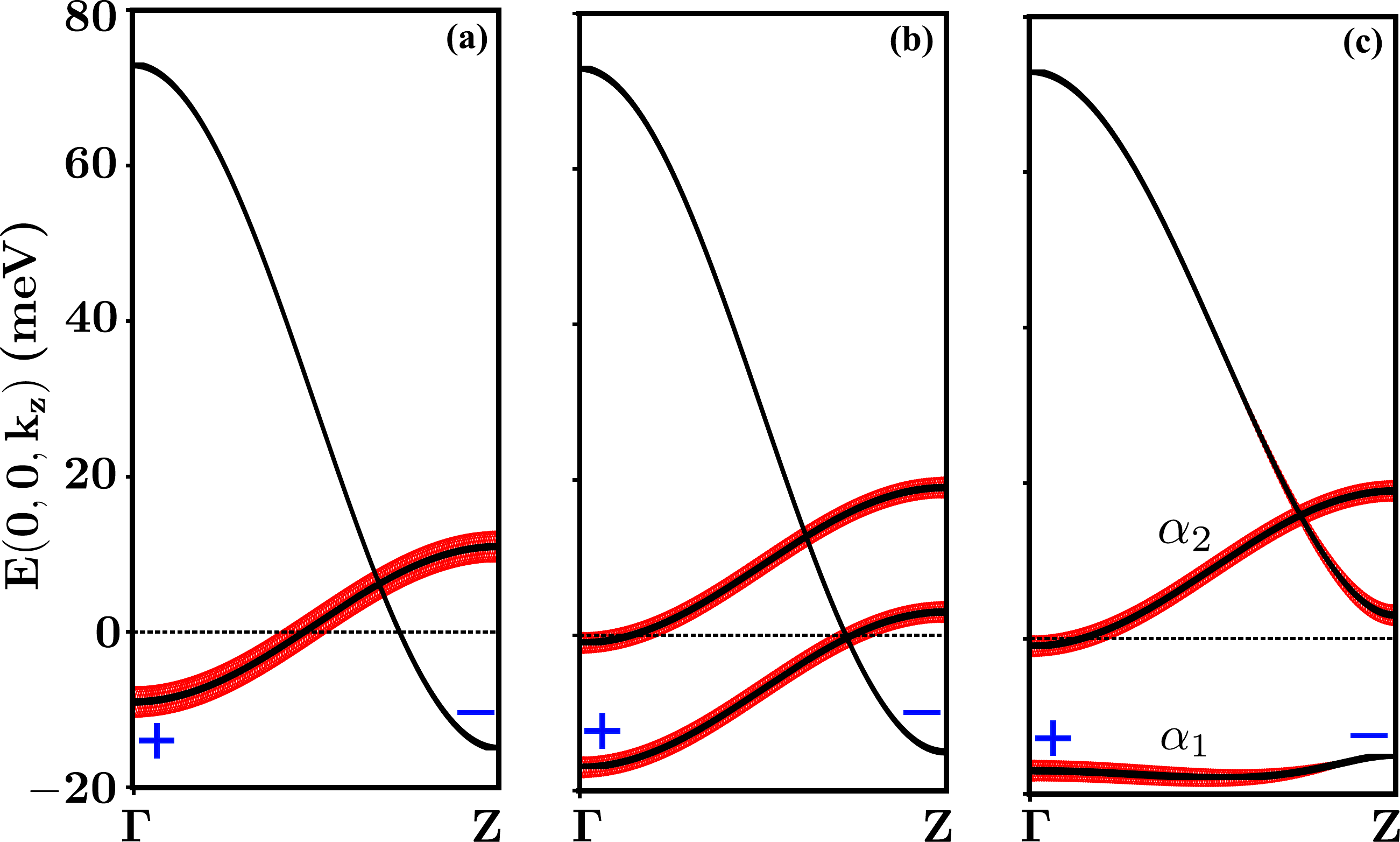}
\caption{\label{fig4}(a) Bulk dispersion along $\Gamma Z$ with all spin-orbit coupling (SOC) turned off (left), with splitting between $d$ orbitals turned on (middle) and with SOC between $p$ and $d$ orbitals turned on (right). The parity of the bands at the time-reversal invariant momenta $\Gamma$ and $Z$ is indicated by the blue symbols. The flat $\alpha_1$ band undergoes band inversion. The size of the markers schematically represents the weight of the $d_{xz}$ orbital, which is visible in LV polarization; see text.}
\end{figure}

We write a minimal model involving $p_z$, $d_{xz}$ and $d_{yz}$ bands motivated by DFT. 
The most general Hamiltonian, which includes up to nearest-neighbor hopping, is 
$H= \sum_{\veck} \Psi_\veck^\dagger h_\veck \Psi_\veck $ with the annihilation operator 
$\Psi_\veck = (\pz, \dyz, \dxz)^T$ and 
the Hermitian matrix
\begin{align}
&h_\veck=
\begin{pmatrix}
\epsilon_p^0  + 2t_{zp}\cos k_z & -2\lambda_3\sx \sin k_z  & 2\lambda_3\sy \sin k_z \\
. & 
\epsilon_d^0 + 2t_{zd}\cos k_z & 
i\sigma_z \left( \lambda_1  + 2\lambda_2 \cos k_z \right)  \\
. & 
. &
\epsilon_d^0 + 2t_{zd}\cos k_z
\end{pmatrix}\label{eq:hamGZ}
\end{align}
where the c-axis lattice spacing $c=1$.
The off-diagonal terms arise from spin-orbit coupling (SOC) and their form is constrained by symmetry.
For instance, to obtain the p-d mixing terms, we note that the even-parity band must 
transform like $p_z$ under the $C_{4v}$ transformations that leave 
$\mathbf{k}=(0,0,k_z)$ invariant. Only then can the two bands hybridize along $\Gamma Z$.
One can check that the operator $\sigma_x c^\dagger_{i,d_{yz}} - \sigma_y c^\dagger_{i,d_{xz}}$ transforms according to the 
trivial representation of $C_{4v}$, just like the $p_z$ orbital. The additional form factor of $\sin k_z$ is required by
inversion symmetry. See Appendix.~\ref{sec:derivation} for details.

Guided by the experimentally observed dispersion in Fig.~\ref{fig3} (a,b),
we choose parameter values (all in  meV) $\epsilon_p^0=29$, 
$\epsilon_d^0=1$, $t_{zp}=22$, $t_{zd}=-5$, and $\lambda_i$'s described below.
The resulting band structure is shown in Fig.~\ref{fig4}. In Fig.~\ref{fig4}(a), where
all $\lambda_i$'s are set to zero, we see the dispersive $p_z$-band and the degenerate $d_{xz}$ and $d_{yz}$ bands.
The ratio $|t_{zp}/t_{zd}|$ is chosen to be similar to the DFT value, although both hoppings are 
suppressed by interactions. In Fig.~\ref{fig4}(b), we see the d-d splitting arising from 
$\lambda_1=8$  meV, keeping $\lambda_2=0$ for simplicity. At this stage, the lower band eigenfunctions are equal admixtures 
of $ d_{xz} $ and $ d_{yz} $ orbitals, i.e. $ \left( |d_{xz}\rangle + i |d_{yz}\rangle \right) |\downarrow\rangle $ and its time-reversed partner.

Finally, in Fig.~\ref{fig4}(c) we turn on p-d mixing $\lambda_3=8$  meV and
obtain an essentially flat $\alpha_1$ band, together with an $\alpha_2$ band that retains its dispersion.
Thus we can understand the ARPES observations with reasonable parameter values for band
renormalizations and SOC. In Appendix~\ref{sec:surface}, we show that these results are obtained for a range of
parameters and not fine-tuned.

We also see from Fig.~\ref{fig4}(c) that the orbital character of the $\alpha_1$ band changes from $d$-like to $p$-like going from $\Gamma$ to $Z$
with a corresponding change in parity eigenvalue. This band inversion is responsible for the non-trivial Fu-Kane invariant~\cite{fu}
of the topological band structure in inversion-symmetric FeSeTe. 

\begin{figure*}
\includegraphics[width=12cm,keepaspectratio]{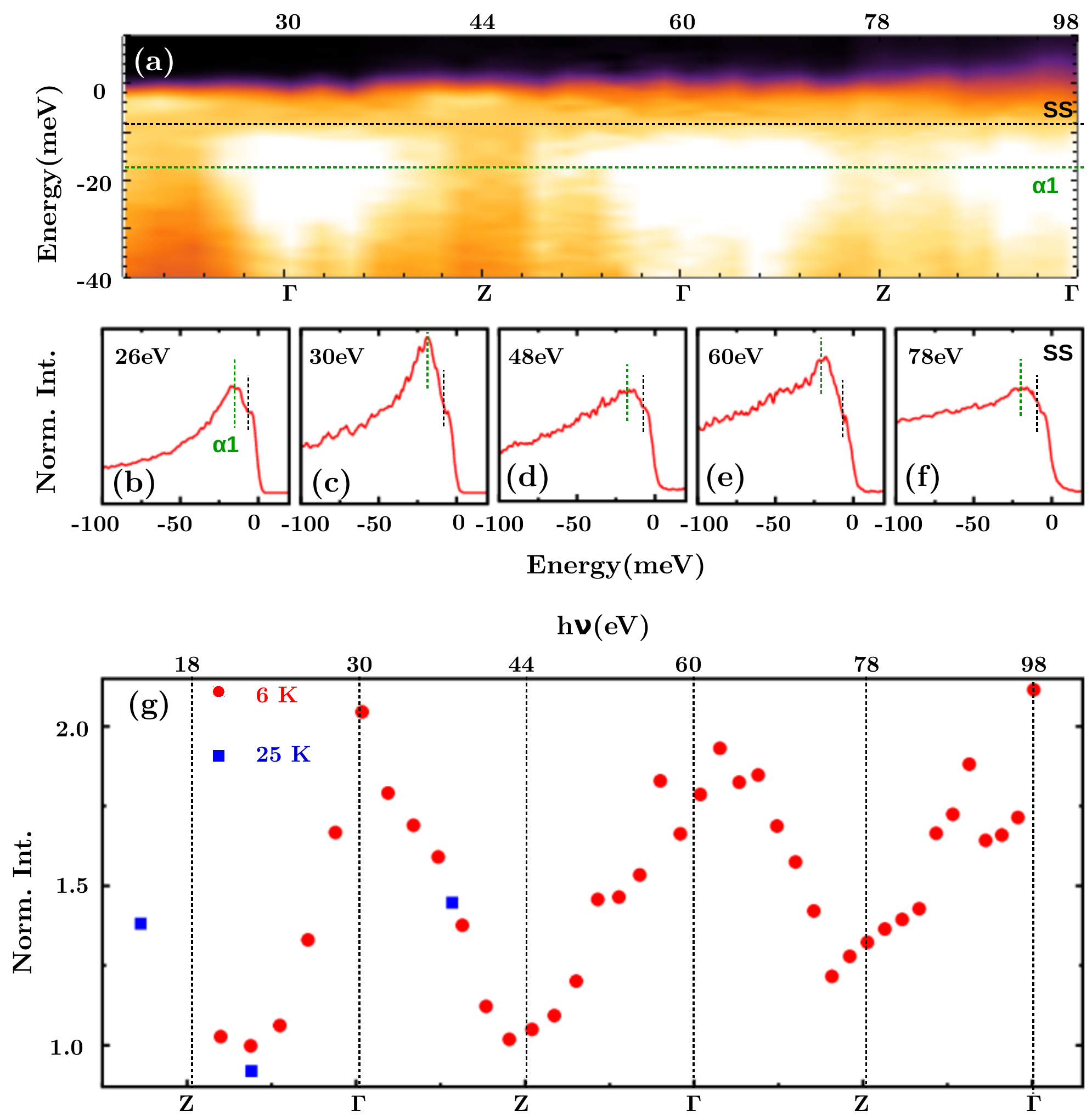}
\caption{\label{fig5} Periodic variation of orbital character of $\alpha_1$
as a signature of band inversion: (a): Band dispersion along the $k_z$ direction which is
prepared  from the EDCs at $k_\parallel = 0$ at different photon energies (20 to 98 eV) using LV polarized light.
Here, EDCs at different photon energies are normalized to have the same spectral weight at the SS position (BE $= 8$  meV).
Green and black dotted lines mark the BE of the $\alpha_1$ and SS respectively.
In (b,c,d,e and f) the normalized EDC for the selected  photon energies 26, 30, 48, 60 and 78 eV are displayed, where
 vertical dashed green and black lines indicate  the position of the $\alpha_1$ and SS respectively.
(g) The normalized ARPES intensity of the $\alpha_1$ band plotted as
a function of incident photon energy, or equivalently $k_z$.
The observed periodic variation of the normalized $\alpha_1$ intensity as a function of $k_z$
with a maximum at $\Gamma$ and a minimum at $Z$ is consistent with the
band inversion characteristic of a topological electronic structure.}
\end{figure*}

We next show how this impacts ARPES selection rules~\cite{herm}
by looking at the matrix element $\langle \psi_f | \mathbf{A \cdot  p} | \psi_i \rangle$ in the experimental geometry of Fig.~\ref{fig1}(a).
For normally emitted photoelectrons, only those final states $| \psi_f \rangle$ that are even under reflections
 in the YZ-plane ($\pi_X$) and  in the XZ-plane ($\pi_Y$) have non-zero amplitude at the detector. 
For LV polarization, $\mathbf{A}\parallel \mathbf{\hat{X}}$ (Fig.~\ref{fig1}(a)) which is odd under $\pi_X$ and even under $\pi_Y$.
This implies only initial states $ | \psi_i \rangle$ which are odd under $\pi_X$ and even under $\pi_Y$ should be visible.
Thus, when lab (XY) and crystal (xy) axes are aligned, as in Fig.~\ref{fig1}(a), only $d_{xz}$ initial states contribute to ARPES with LV polarization. 
(see Appendix~\ref{sec:selection}  for a detailed review of selection rules in other cases.)

In Fig.~\ref{fig4}(c), we schematically indicate by the width of the red line the weight of the $d_{xz}$ orbital.
We thus expect the ARPES intensity in LV polarization to exhibit strong variation with photon energy, as $k_z$
varies from $\Gamma$, where you expect a strong contribution from the even parity $d_{xz}$ state, to $Z$,
where the intensity from $d_{xz}$ should be suppressed.

To test this experimentally, we need to normalize the ARPES intensity before we can compare the signals at two different photon
energies. A convenient normalization is to use the $k_z$-independent surface state (SS) at BE $= 8$  meV discussed above (see Fig.~\ref{fig2}).
In Fig.~\ref{fig5}(g), we plot the variation with photon energy of the intensity at the top of the $\alpha_1$ band in LV polarization, normalized at each photon energy by the intensity at BE $=8$  meV. 
We thus obtain the main result of our paper: a clear periodic variation of the normalized intensity over a range of photon energy spanning almost three Brillouin zones,
 with a maximum at $\Gamma$ and a minimum at $Z$, fully consistent with the band inversion characteristic of a topologically non-trivial bulk band structure.

We note that this periodic variation in the normalized intensity at the top of $\alpha_1$ is clearly visible in the EDC of normally emitted photoelectrons shown for a few selected photon energies in Fig.~\ref{fig5}(b-f), and in the variation of intensity with photon energy on the x-axis and binding energy on the y-axes in Fig.~\ref{fig5}(a).

In conclusion, the electronic structure of FeSeTe poses a unique challenge where one has to deal with both 
strong correlations and band topology. 
DFT calculations predict a highly dispersive bulk band structure along $k_z$ which seems to be essential
for the band inversion that leads to a topologically protected Dirac surface state.
However, the observed bulk band structure is strongly renormalized by correlations and shows essentially no $k_z$-dispersion, 
in marked contrast with the DFT predictions, and raises doubts about band inversion.
Through a combination of extensive ARPES data as a function of photon energy and
a careful examination of the orbital character of the bulk bands using selection rules, we
show that the $\alpha_1$ band at the $Z$ point is in fact inverted with respect to $\Gamma$,
despite the lack of $k_z$-dispersion. 
Our modeling provides a natural explanation in terms of renormalized band widths 
that are comparable to the spin-orbit coupling.
We thus reveal an unusual situation where an almost flat band undergoes band inversion,
characteristic of a topologically non-trivial bulk band structure.
  
{\bf Methods:}

{\it Sample preparation -} High-quality single crystals of Fe$_{1.02}$Se$_{0.45}$Te$_{0.55}$ were grown using the modified Bridgman method. The stoichiometric amounts of high-purity Fe, Se, and Te powders were grinded, mixed, and sealed in a fused silica ampoule. The ampoule was evacuated to a vacuum better than 10$^{-5}$ torr, and the mixture was reacted at $750^\circ$C for 72 hours. The resulting sinter was then regrinded and put in a double-wall ampoule that was again evacuated to a vacuum better than 10$^{-5}$ torr.

The ampoule was placed in a two-zone furnace with a gradient of $5^\circ$C/cm and slowly cooled from $1040^\circ$C to $600^\circ$C at a rate of $2^\circ $C/hour, followed by a faster cooldown to $360^\circ $C for 24 hours. The resulting boule contained single crystals that could be separated mechanically. To improve the uniformity of the superconducting phase, we annealed the crystals for 48 hours in ampoules that were evacuated and then filled them with 10$^{-3}$ torr  of oxygen.  Crystallinity of the prepared single crystals confirmed  by XRD measurements and elemental composition determined through energy dispersive X-ray (EDX) analysis~\cite{kanigel1,kanigel}.

{\it ARPES - } High-resolution ARPES measurements were performed at the UE112-PGM-2b-1$^3$ beamline at BESSY (Berlin, Germany), at the I05 beamiline at Diamond (Didcot, UK) and at the SIS beamline at the SLS, PSI (Villigen, Switzerland) using photon energies between 15 eV and 150eV. The samples were cleaved in vacuum better than 5 $\times$10$^{-11}$ torr at low temperature and measured for not more than 6 hours. The base temperature at BESSY was 1 K and at Diamond was 6 K. The energy resolution was 4 meV in these beamlines. At PSI, the temperature was 25 K and and the energy resolution was 10 meV. 

{\it DFT - }To resolve the band structure of FeSe$_{0.45}$Te$_{0.55}$, density functional theory (DFT) calculations with spin-orbit coupling were performed using the Vienna ab initio simulation package (VASP) with core electrons represented by the projector-augmented-wave (PAW) potential~\cite{DFT1}. Generalized-gradient-approximation (GGA)~\cite{DFT2}  functional was used for the exchange-correlation potential. To treat the alloy, we perform the DFT calculation using the virtual crystal approximation with ordered Se and Te sites in the two-formula cell. Plane waves with a kinetic energy cutoff of 300 eV were used as the basis set. A k-point grid of $20\times20\times20$ was used for Brillouin zone sampling. 

{\bf Acknowledgements}: M.R. would like to thank A. Chubukov and H. Ding for stimulating discussions.
We gratefully acknowledge support from the US-Israel Binational Science
Foundation grant 2014077. H.L. is supported in part at Israel Institute of Technology, Technion by a  PBC fellowship of the Israel Council for Higher Education.
Work at the Technion was supported by the Israeli Science Foundation Grant 320/17. 
We thank the Helmholtz-Zentrum Berlin for the allocation of synchrotron radiation beamtime. 
We acknowledge Diamond Light Source for time on Beamline I05 under Proposal SI15822.
We acknowledge the Paul Scherrer Institut, Villigen, Switzerland for provision of synchrotron radiation beamtime at beamline SIS of the SLS.
The research leading to these results has received funding from the European Union's Horizon 2020 research and innovation programme under grant agreement no 730872, project CALIPSOplus

\appendix

\section{Photon energy to $k_z$ mapping}\label{sec:kzmapping}

 The relation $k_z$ = $\sqrt{\frac{2m}{\hbar^2}}\sqrt{(h\nu-\phi-E_B)cos^2\theta+V_0}$, where $\phi$, E$_B$, $\theta$ and V$_0$ correspond to work function, BE of photoelectron, emission angle of photoelectron with respect to the sample normal and  inner potential of the sample respectively\cite{huf}, allows a conversion between incident photon energy and $k_z$ for a known inner potential. 
 
 The constant V$_0$ is specific to the material and can be determined experimentally from the ARPES data, by identifying the high-symmetry points in the dispersion along $k_z$ of a band. 

\begin{figure*}
\includegraphics[width=13cm,keepaspectratio]{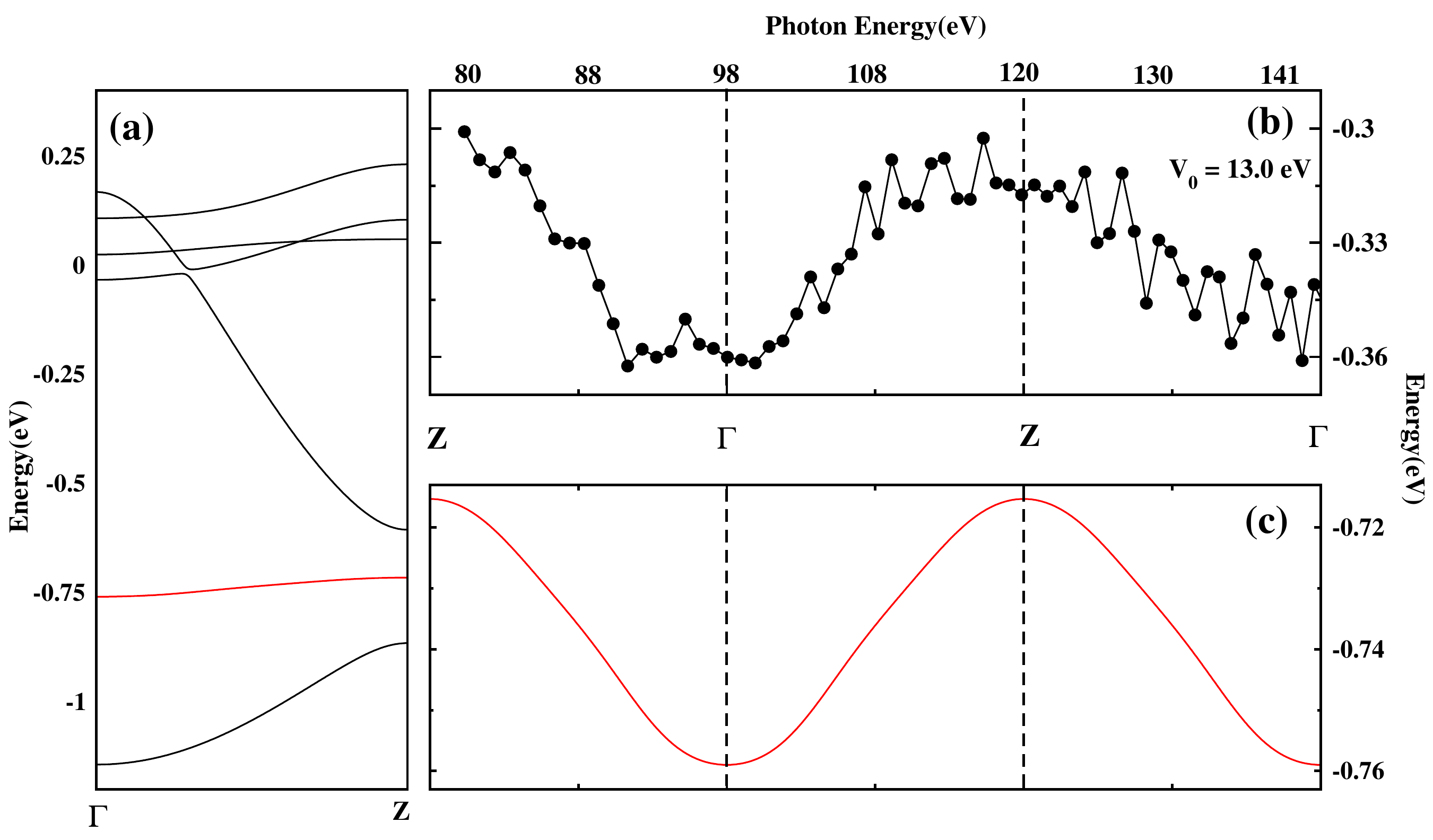}
\caption{\label{fig6}(a) Band structure of FeSeTe along $\Gamma$-Z direction calculated from the DFT. (b) Experimental band  of FeSeTe with respect to photon energy(upper x-axis)
and  $k_z$/($\pi$/c)(lower x-axis) observed in ARPES measurements. Here, $\pi/c$ is reciprocal lattice vector corresponding to the lattice parameter c of FeSeTe.
 The experimental band  exhibits a high qualitative  resemblance to the DFT band in Fig.\ref{fig6}(a) marked with red colour. This calculated band is plotted separately 
an extended Brillouin zone in (d). In Fig.\ref{fig6}(b) inner potential V$_0$ = 13.0  eV is used to convert the photon energies dependency  of the experimental band   to corresponding $k_z$
dispersion.  
}
\end{figure*}

Fig.\ref{fig6}(a) shows FeSeTe  band dispersion along the $\Gamma$(0,0,0) to Z(0,0,$\pi$/c) direction calculated using DFT. We choose to use the d$_{z^2}$ band, marked in red, for extracting the value of the inner potential.
For that purpose we measured the ARPES spectra at normal emission over a large binding energy range for  photon energies varying between 80 eV to 150 eV. The measurements were done at 25K in a p-type configuration, where the incident 
plane of light is parallel to the analyser slit with the sample oriented along the $\Gamma$-M direction. 

The results, binding energy of d$_{z^2}$ as a function of photon energy,  are shown in Fig.\ref{fig6}(b).
The DFT dispersion of the same band as a function of $k_z$ is shown in Fig.\ref{fig6}(c). 
The best agreement between the experimental and calculated bands is achieved for an inner potential value of 13 eV.  The measured bandwidth is in reasonable agreement with the calculation although the band is shifted by about 400 meV. 
 
\section{ARPES Selection Rules}\label{sec:selection}

\begin{figure*}
\includegraphics[width=12cm,keepaspectratio]{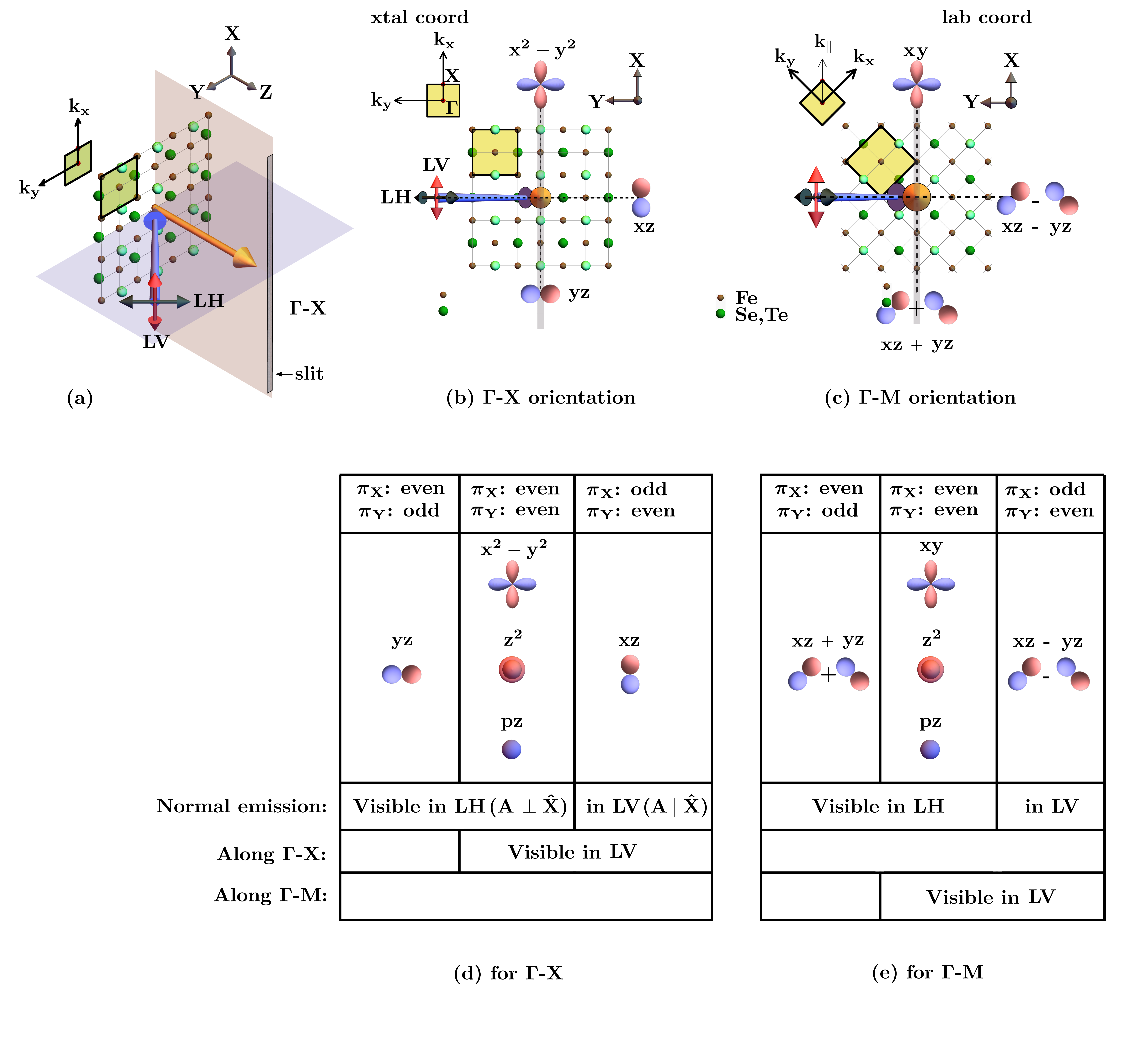}
\caption{\label{figS1} (a) Schematic view of s-type ARPES set-up, with incident plane (blue) horizontal (YZ), and emission plane (orange) vertical (XZ).
We use $(X,Y,Z)$ to denote lab frame coordinates and $(x,y,z)$ for sample coordinates consistent with the crystal axes.
(b,c): Views along Z-axis of the sample oriented with analyzer slit along $\Gamma$-X (in panel (b)) and oriented along $\Gamma$-M (in panel (b)).
We also show orbitals that are eigenstates of reflection in the incident and emission planes. (d and e): ARPES selection rules corresponding to panels
 (b) and (c) respectively; see text for details. 
}
\end{figure*}

We now review the ARPES selection rules for the s-configuration setup shown schematically in Fig.\ref{figS1}(a)
 by considering the matrix element $\langle \psi_f | \mathbf{A \cdot  p} | \psi_i \rangle$.
Detection of the final state requires that its wavefunction $|\psi_f \rangle $ must be invariant under the symmetries that keep the emitted ray (orange arrow) invariant. For normal emission, this includes both $\pi_Y$: reflection about the emission plane (XZ), and $\pi_X$: reflection about the incident plane (YZ). Away from normal emission, only $\pi_Y$ keeps the emitted ray invariant. If the polarization vector $\mathbf{A}$ has definite parity under these symmetries, the matrix element is non-zero only for orbitals that have the same parity. Only those bands are visible which have finite weight in these orbitals. The selection rules for different cases are explicitly shown in Fig.~\ref{figS1}(d). Note that when the sample is rotated as in Fig.~\ref{figS1}(c), the orbitals with definite parity under reflection are not $d_{xz},d_{yz}$ defined with reference to the crystallographic axes~\cite{wang}. For LH polarization away from normal emission, there are no symmetry-enforced selection rules since the polarization vector does not have definite parity under $\pi_Y$.

\begin{table}
\caption{\label{table} List of allowed orbitals at normal emission for LH and LV polarization.}
\begin{tabular}{cc}
\hline

\hline
\multicolumn{2}{c}{Sample orientation along $\Gamma$-X} \\
\hline
\hline
Polarization&Allowed Orbital\\
\hline
LH&d$_{yz}$,d$_{x^2-y^2}$,d$_{z^2}$,p$_z$\\
LV&d$_{xz}$\\
\hline
\multicolumn{2}{c}{Sample orientation along $\Gamma$-M} \\
\hline
\hline
LH&(d$_{xz}$+d$_{yz}$),d$_{xy}$,d$_{z^2}$,p$_z$\\
LV&(d$_{xz}$-d$_{yz}$)\\
\hline
\end{tabular}
\end{table}

\section{Estimation of top of $\alpha_2$ band}\label{sec:alpha2}
Fig.\ref{fig4}(a) represents ARPES image of FeSeTe sample along the $\Gamma$-M direction which is collected  at photon energy 22  eV 
utilizing photons of LH polarization.  MDC line profile at BE = 25 meV  extracted from  this image is
displayed in Fig.\ref{fig4}(b). This MDC profile is fitted to Lorentzian function with three peaks to
track the band dispersion of the $\alpha$2 band  as shown in Fig.\ref{fig4}(c). This band dispersion is 
fitted to  a parabolic dispersion to estimate the apex of  $\alpha$2($\epsilon_{\alpha 2}$) band.

\begin{figure}
\includegraphics[width=8cm,keepaspectratio]{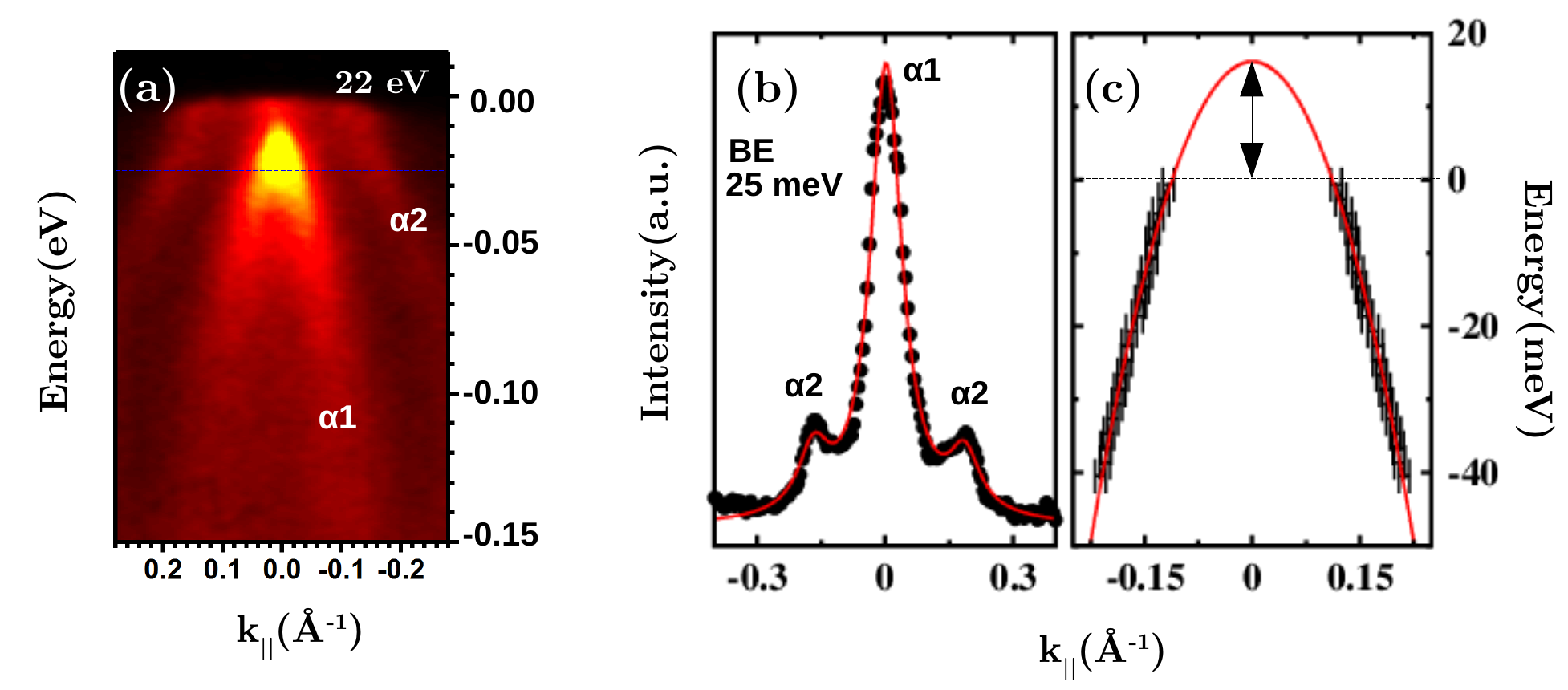}
\caption{\label{figS4} (a) ARPES image of sample FeSeTe oriented along the $\Gamma$-M direction at
  photon energy of 22  eV using LH polarization and at temperature  $\sim$ 6 K. (b) MDC cut at
BE = 25 meV(blue dotted line) from the ARPES image (a), where red curve is Lorentzian function fitted  to this MDC data.  
(c) Band dispersion of $\alpha$2 band obtained from the MDC peak fitting(b) and its top($\epsilon_{\alpha 2}$)
  is estimated from the parabolic fitting(red). }
\end{figure}

\section{Derivation of model Hamiltonian}\label{sec:derivation}
The space group of FeSe$_{0.45}$Te$_{0.55}$ is actually {\it P4/nmm} because
of the buckled square lattice, with the chalcogen atoms alternately above and below the Fe plane. We use the existence of a
one-to-one mapping~\cite{cve} between the point group {\it D$_{4h}$} and the
space group {\it P4/nmm} modulo lattice translations, to describe
the physics near $\Gamma$-Z in terms of the more well-known {\it D$_{4h}$} group.

We use the method of invariants~\cite{lut} to construct the most general symmetry-allowed Hamiltonian $H=\sum_n c_n I_n$ where $I_n$ are fermionic bilinears of the form
\begin{align}
I_n = \sum_{\mathbf{k},i\mu\nu\alpha\beta} h^i_{n\mu\nu}(\mathbf{k}) \, c_{\mathbf{k}\mu\alpha}^\dagger   \sigma^i_{\alpha\beta} c_{\mathbf{k}\nu\beta}
\end{align}
which are invariant under time-reversal and the $D_{4h}$ symmetry group.
Here $c_{\mathbf{k}\mu\alpha}^\dagger$ creates an electron with crystal momentum $k$ in the orbital $\mu\in \{d_{xz},d_{yz},p_z\}$ with spin $\alpha$ and $i=0..4$.
For simplicity, we henceforth use the notation, $c_{\mathbf{k}d_{xz}}^\dagger \equiv d_{xz}^\dagger$, etc. 
Since we are interested in a minimal description of the physics in the vicinity of $\Gamma$-$ Z $, we consider terms upto quadratic order in $k_x$ and $k_y$ and upto nearest neighbor hopping along $z$. The symmetry constrained Hamiltonian takes the form $H= \sum_{\veck} \Psi_\veck^\dagger \tilde{h}_\veck \Psi_\veck $ with $\Psi_\veck = (\pZ, \dYZ, \dXZ)^T$ and the Hermitian matrix  
\begin{align}
&\tilde{h}_\veck=
\begin{pmatrix}
\epsilon_p & f_y(\veck) & f_x(\veck) \\
\cdot & 
\epsilon_d + (k_x^2-k_y^2)\mathpzc{C}_1 & 
f_{xy}(\veck) \\
\cdot & 
\cdot &
\epsilon_d - (k_x^2-k_y^2)\mathpzc{C}_1 
\end{pmatrix} \notag\\
&{\rm with}\qquad \notag\\
&
f_{x}(\veck)= \left[\mathpzc{S}_0 +  (k_x^2 +k_y^2) \mathpzc{S}_1 + (k_x^2 -k_y^2) \mathpzc{S}_2 \right]\sy +k_x k_y \sx \mathpzc{S}_3 \notag\\
&\qquad\qquad\qquad\qquad+ i k_x \mathpzc{C}_2 -k_y \sz \mathpzc{C}_3 \notag\\
&
f_{y}(\veck)=-\left[\mathpzc{S}_0 +  (k_x^2 +k_y^2) \mathpzc{S}_1 - (k_x^2 -k_y^2) \mathpzc{S}_2 \right]\sx -k_x k_y \sy \mathpzc{S}_3 \notag\\
&\qquad\qquad\qquad\qquad+ i k_y \mathpzc{C}_2 +k_x \sz \mathpzc{C}_3  \notag\\
&
f_{xy}(\veck)=k_x k_y \mathpzc{C}_4 + i\sz \epsilon_{\rm SOC} + i (k_\parallel. \sigma_\parallel) \mathpzc{S}_4 
\label{eq:modelHam}
\end{align}
where $\epsilon_\mu=\epsilon_\mu^0  + k_\parallel^2/2m_\mu + t_{z\mu}\cos k_z + t_{\rm diag\mu}k_\parallel^2\cos k_z $, $\mathpzc{C}_i=t_{i0}+t_{iz}\cos k_z$ and $\mathpzc{S}_i=t'_{i}\sin k_z$ are even and odd functions of $k_z$, $k_\parallel$ and $\sigma_\parallel$ are the in-plane projections of $\veck$ and $\boldsymbol{\sigma}$, all coefficients are real and all momenta are rescaled by the appropriate lattice constants $k_x\rightarrow k_x\,a$, $k_y\rightarrow k_y\,a$, $k_z \rightarrow k_z\, c$. In the following, we step through the derivation of this Hamiltonian by enumerating the eigenstates of the $D_{4h}$ point group symmetries.

$D_{4h}$ is generated by $C_4$, $\pi_x$ (reflections in the YZ plane) and inversion. 
Under $C_4$, $d_{xz} \rightarrow d_{yz}, \, d_{yz} \rightarrow -d_{xz}$, and we list below their Hermitian bilinears that are eigenstates of $C_4$
\begin{table}[h]
\centering
\begin{tabular}{|c|c|c|}
\hline
  & $C_4$ & $\pi_x$ \\
\hline
$\dXZ^\dagger$ $\dXZ$ + $\dYZ^\dagger$ $\dYZ$ & 1 & 1 \\
$\dXZ^\dagger$ $\dXZ$ - $\dYZ^\dagger$ $\dYZ$ & -1 & 1 \\
$\dXZ^\dagger$ $\dYZ$ + $\dYZ^\dagger$ $\dXZ$ & -1 & -1 \\
$i\dXZ^\dagger$ $\dYZ$ - $i\dYZ^\dagger$ $\dXZ$ & 1 & -1 \\
\hline
\end{tabular}
\end{table}.
Clearly, $\dXZ^\dagger \dXZ + \dYZ^\dagger \dYZ$ is invariant under the point group symmetries, and $i(\dXZ^\dagger \dYZ - \dYZ^\dagger \dXZ )$ transforms like $\sigma_z$. The most general on-site Hamiltonian is therefore
\begin{align}
h_{\veck0} = \epsilon_p^0 \pZ^\dagger \pZ +\epsilon_d^0 \left(\dXZ^\dagger \dXZ + \dYZ^\dagger \dYZ\right) + i\sigma_z \lambda_1 \left( \dXZ^\dagger \dYZ - \dYZ^\dagger \dXZ  \right)\label{eq:kp1}
\end{align}
where the coefficients are required to be real because of time-reversal invariance. 

Next we consider nearest neighbor hopping along $z$. The availability of an inversion-odd form factor allows the possibility of hybridization between $p_z$ and an appropriate combination of $d$ orbitals that is invariant under $C_4$ and $\pi_x$. Such a combination results from mixing with the in-plane spin operators which also transform into each other under $C_4$
\begin{table}[h]
\centering
\begin{tabular}{|c|c|c|}
\hline
 & $C_4$ & $\pi_x$ \\
\hline
$\sigma_x \dYZ - \sigma_y \dXZ$ & 1 & 1 \\
$\sigma_x \dYZ + \sigma_y \dXZ$ & -1 & 1 \\
$\sigma_x \dXZ - \sigma_y \dYZ$ & -1 & -1 \\
$\sigma_x \dXZ + \sigma_y \dYZ$ & 1 & -1 \\
\hline
\end{tabular}
\end{table}. 
The Hamiltonian along $\Gamma-Z$ thus has the following extra terms with the out-of-plane momentum $k_z \rightarrow k_z c$ rescaled by the c-axis lattice constant $c$
\begin{align}
h_{\veck z}=&2\cos k_z \big[ t_{zp}^0 \pZ^\dagger \pZ +t_{zd}^0 \left(\dXZ^\dagger \dXZ + \dYZ^\dagger \dYZ\right) + \nonumber\\
						&\qquad\qquad\qquad i\sigma_z \lambda_2 \left( \dXZ^\dagger \dYZ - \dYZ^\dagger \dXZ  \right) \big] \nonumber\\
		   &-2\sin k_z \lambda_3 \left[ \pZ^\dagger \left( \sigma_x \dYZ - \sigma_y \dXZ \right) + \mathrm{h.c.} \right]
		   \label{eq:kp2}
\end{align}
where time-reversal invariance again requires the coefficients to be real. This leads to the model in Eq.~
\eqref{eq:hamGZ}.

The in-plane dispersion in the vicinity  of $\Gamma Z$ can be modelled by $\mathbf{k\cdot p}$ perturbation theory, neglecting terms cubic or higher order in $k_\parallel$.
At quadratic order, the $C_4$-odd form factors $k_x^2-k_y^2$ and $k_x k_y$ result in the following terms
\begin{align}
h_{\veck\parallel,1}= \left( \frac{k_x^2-k_y^2}{2m'} \right) &\left( \dXZ^\dagger \dXZ - \dYZ^\dagger \dYZ \right) +
						\frac{k_x k_y}{2m''} \left( \dXZ^\dagger \dYZ + \dYZ^\dagger \dXZ \right)  \nonumber \\
					 +\left( \frac{k_x^2-k_y^2}{2m_{\rm SOC,1}} \right) \pZ^\dagger &\left( \sigma_x \dYZ + \sigma_y \dXZ \right) +
					    \frac{k_x k_y}{2m_{\rm SOC,2}} \pZ^\dagger \left( \sigma_x \dXZ - \sigma_y \dYZ \right)\label{eq:kp3}
\end{align}
in addition to those derived from each of the invariants in Eqs.~\eqref{eq:kp1},\eqref{eq:kp2} by multiplying the $C_4$ invariant form factor $k_x^2 + k_y^2$ and in Eq.~\eqref{eq:kp3} by multiplying the inversion-even form factor $\cos k_z$. Here and henceforth, $k_\parallel$ is rescaled by the in-plane lattice constant $a$: $k_{x,y} \rightarrow k_{x,y} a$ and $\hbar=1$.

In addition, there are linear terms in $k_\parallel$ combined with $\sigma_\parallel$ into the following $C_4$ eigenstates.
\begin{table}[h]
\centering
\begin{tabular}{|c|c|c|}
\hline
  & $C_4$ & $\pi_x$ \\
\hline
$\sigma_x k_y - \sigma_y k_x$ & 1 & 1 \\
$\sigma_x k_y + \sigma_y k_x$ & -1 & 1 \\
$\sigma_x k_x - \sigma_y k_y$ & -1 & -1 \\
$\sigma_x k_x + \sigma_y k_y$ & 1 & -1 \\
\hline
\end{tabular}
\end{table}
This results in only one additional term in the Hamiltonian
\begin{align}
h_{\veck,\times}=-i v_\times \sin k_z \left( \sigma_x k_x + \sigma_y k_y \right) \left( \dXZ^\dagger \dYZ - \dYZ^\dagger \dXZ  \right)
\end{align}
the other three similar terms being inadmissible since hermiticity requires their coefficients to be real and time-reversal invariance requires them to be imaginary.

Lastly, away from $\Gamma Z$, $k_\parallel$ combines with $d$-orbitals to allow additional $p-d$ coupling terms of the form
\begin{align}
h_{\veck,pd}=&\left(v_{pd0} + 2v_{pdz} \cos k_z \right) \pZ^\dagger \left( k_x \dXZ +k_y \dYZ \right) +\mathrm{h.c.}\nonumber\\
			+&\left(v'_{pd0} + 2v'_{pdz} \cos k_z \right) \sigma_z \pZ^\dagger \left( k_x \dYZ - k_y \dXZ \right) +\mathrm{h.c.} .
\end{align} 
This leads to the most general $\mathbf{k \cdot p}$ Hamiltonian consistent with the point group and time-reversal symmetries, which has the form shown in Eq.~\eqref{eq:modelHam}.

\begin{figure}
\includegraphics[width=0.25\textwidth]{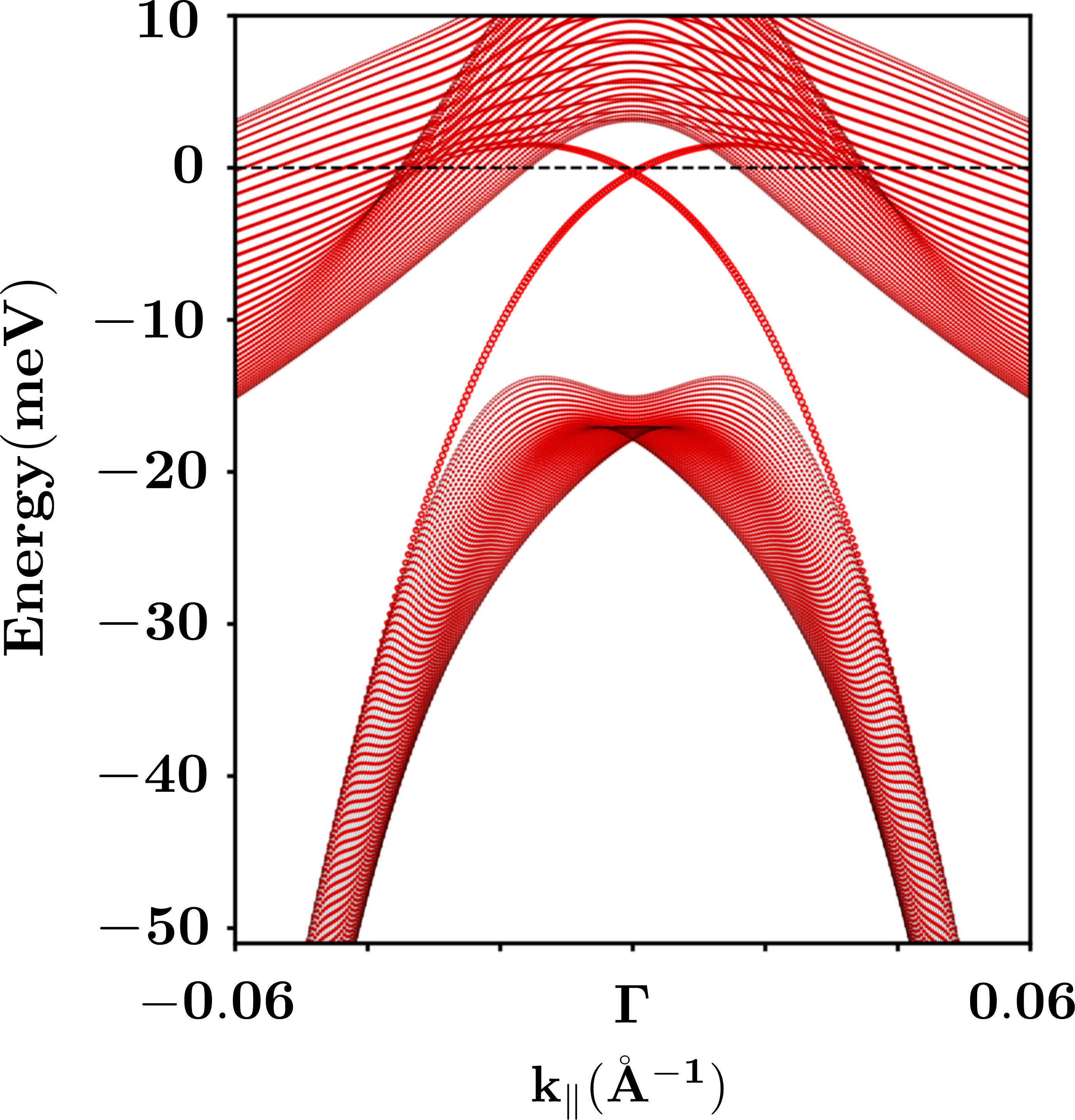}
\caption{\label{fig:dirac} Dirac surface state as seen in the calculated bandstructure of a 40-layer slab. Marker size is indicative of the weight on the top layer. Here, $\epsilon_p^0=29$, $\epsilon_d^0=1$, $t_{zp}=22$, $t_{zd}=-5$, $ \lambda_1 =8 $, $ \lambda_3 =8 $, $ 1/m_p = 3000$, $ 1/m_d =-1500 $, $ 1/m''=-5700 $, $ v_{pd0} =82 $ (in meV) and all other parameters are set to 0.
}
\end{figure}

\begin{figure}
\includegraphics[width=0.45\textwidth]{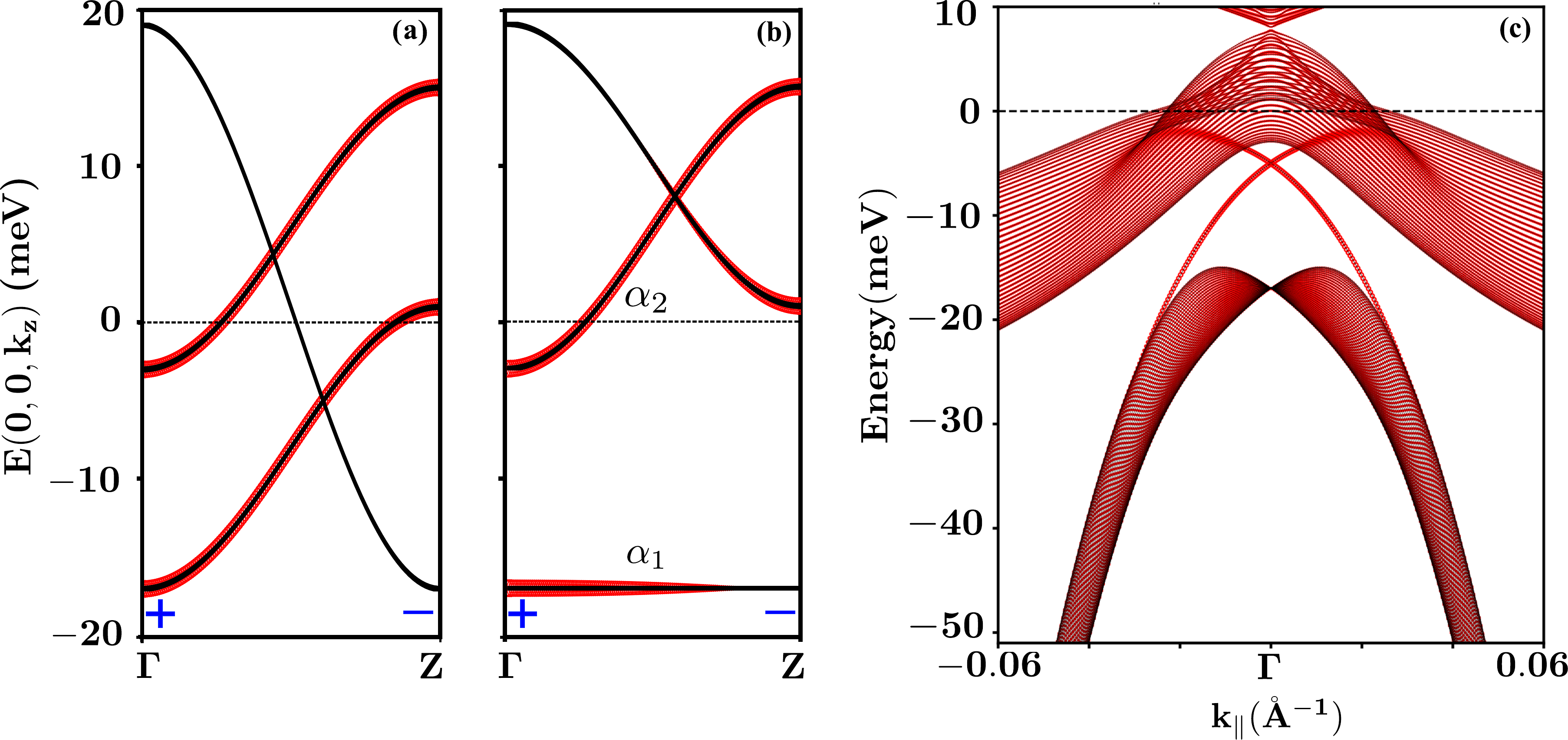}
\caption{\label{fig:model2} Model calculations for a different choice of parameters, showing (a) dispersion along $\Gamma Z$ without $ d $-$ p $ mixing $ \epsilon_p^0=1 $, $ \epsilon_d^0 =-1 $, $ t_{zp}=9 $, $ t_{zd} =-4.5 $, $ \lambda_1=7 $, (in meV) (b) with $p-d$ mixing turned on, $\lambda_3=4.5$, (c) surface bandstructure of 40-layer slab with $ 1/m_p = 3000$, $ 1/m_d =-1500 $, $ 1/m''=-5700 $, $ v_{pd0} =82 $ as before.
}
\end{figure}

\section{Calculation of surface band structure}\label{sec:surface}

The band inversion along $\Gamma Z$ leads to a protected Dirac cone in the bandstructure of the (001) surface. To see this, we must consider a model in a slab geometry with translation symmetry broken along $z$. For simplicity, we consider the Hamiltonian $H=\sum_\veck \Psi_\veck^\dagger \tilde{h}_\veck \Psi_\veck$ along the $ \Gamma M$ direction where $\veck=(k/\sqrt{2},k/\sqrt{2},k_z)$ with 
\begin{align}
\tilde{h}_\veck = h_\veck +
\begin{pmatrix}
\frac{k^2}{2m_p} & i\, v_{pd0} \frac{k}{\sqrt{2}} & i\, v_{pd0} \frac{k}{\sqrt{2}} \\
\cdot & \frac{k^2}{2m_d} & \frac{k^2}{4m''} \\
\cdot & \cdot & \frac{k^2}{2m_d} 
\end{pmatrix}
\end{align}
where $\Psi_\veck = (\pz, \dyz, \dxz)^T$ and $h_\veck$ which describes the physics along $\Gamma Z$ is given by Eq.~
\eqref{eq:hamGZ}. 
Substituting
\begin{align}
\sum_{k_z} 2 \cos k_z \, \, &c_{\veck\mu\sigma}^\dagger c_{\veck\mu'\sigma'} \nonumber \\
 &\rightarrow \sum_{n_z} \left[ c_{\veck_\parallel\mu\sigma}^\dagger (n_z) c_{\veck_\parallel\mu'\sigma'}(n_z+1) + {\rm h. c.} \right] \nonumber \\
\sum_{k_z} 2 \sin k_z \, \, &c_{\veck\mu\sigma}^\dagger c_{\veck\mu'\sigma'} \nonumber \\
 &\rightarrow \sum_{n_z} \left[ic_{\veck_\parallel\mu\sigma}^\dagger(n_z) c_{\veck_\parallel\mu'\sigma'}(n_z+1) + {\rm h. c.} \right]
\end{align}
captures the breaking of translation symmetry at the surface, where $c_{\veck\mu\sigma}^\dagger(n_z)$ creates an electron in orbital $\mu$ with spin $\sigma$ on the layer $n_z$ with in-plane momentum $\veck_\parallel$. The resulting surface bandstructure with the Dirac cone is shown in Fig.~\ref{fig:dirac}.

We note that unlike the $k_z$ dispersion presented in the main text, the model parameters controlling the in-plane dispersion are not constrained by the experimental data.
From the many symmetry-allowed terms (see above) in the in-plane Hamiltonian, we have chosen to retain a minimal set of non-zero parameters that captures the $\alpha_1$ dispersion and the surface Dirac cone. However, the presence of the surface Dirac cone is quite generic, independent of the precise values of the model parameters, as we show in Fig.~\ref{fig:model2} for a slightly different choice of parameters.

\end{document}